\documentclass[acmsmall,screen,nonacm]{acmart}

\usepackage{listings}
\definecolor{halfgray}{gray}{0.35}
\definecolor{deepblue}{rgb}{0,0,0.5}
\definecolor{deepred}{rgb}{0.6,0,0}
\definecolor{deepgreen}{rgb}{0,0.5,0}
\definecolor{highlightorange}{rgb}{0.98, 0.92, 0.8}

\makeatletter
\newcommand\HL{%
	\gdef\lst@alloverstyle##1{%
		\fboxrule=0pt
		\fboxsep=0pt
		\colorbox{lightgray}{\strut##1}%
	}%
}
\newcommand\HLoff{%
	\xdef\lst@alloverstyle##1{##1}%
}

\definecolor{lightgreen}{rgb}{0.9,1,0.9}
\definecolor{lightred}{rgb}{1,0.9,0.9}

\makeatletter
\newcommand\HLgreen{%
	\gdef\lst@alloverstyle##1{%
		\fboxrule=0pt
		\fboxsep=0pt
		\colorbox{lightgreen}{\strut##1}%
	}%
}
\newcommand\HLgreenoff{%
	\xdef\lst@alloverstyle##1{##1}%
}

\makeatletter
\newcommand\HLred{%
	\gdef\lst@alloverstyle##1{%
		\fboxrule=0pt
		\fboxsep=0pt
		\colorbox{lightred}{\strut##1}%
	}%
}
\newcommand\HLredoff{%
	\xdef\lst@alloverstyle##1{##1}%
}

\usepackage{enumitem}
\setlist[itemize]{leftmargin=1.2em}

\newcommand{\code}[1]{\texttt{\small #1}}
\newcommand{\scode}[1]{\texttt{\footnotesize #1}}

\begin{document}

\newcommand{\name}{AgentStepper}

\title{\name{}: Interactive Debugging of Software Development Agents}

\author{Robert Hutter}
\email{research@robert-hutter.de}
\affiliation{%
  \institution{University of Stuttgart}
  \city{Stuttgart}
  \country{Germany}
}

\author{Michael Pradel}
\email{michael@binaervarianz.de}
\affiliation{%
  \institution{CISPA Helmholtz Center for Information Security}
  \city{Stuttgart}
  \country{Germany}}

\begin{abstract}
Software development agents powered by large language models (LLMs) have shown great promise in automating tasks like environment setup, issue solving, and program repair.
Unfortunately, understanding and debugging such agents remain challenging due to their complex and dynamic nature.
Developers must reason about trajectories of LLM queries, tool calls, and code modifications, but current techniques reveal little of this intermediate process in a comprehensible format.
The key insight of this paper is that debugging software development agents shares many similarities with conventional debugging of software programs, yet requires a higher level of abstraction that raises the level from low-level implementation details to high-level agent actions.
Drawing on this insight, we introduce \name{}, the first interactive debugger for LLM-based software engineering agents.
By adapting established debugging practices to agents, \name{} enables developers to inspect, control, and interactively manipulate agent trajectories.
\name{} represents trajectories as structured conversations among an LLM, the agent program, and tools.
It supports breakpoints, stepwise execution, and live editing of prompts and tool invocations, while capturing and displaying intermediate repository-level code changes.
Our evaluation applies \name{} to three state-of-the-art software development agents, ExecutionAgent, SWE-Agent, and RepairAgent, showing that integrating the approach into existing agents requires minor code changes (39--42 edited lines).
Moreover, we report on a user study with twelve participants, indicating that \name{} improves the ability of participants to interpret trajectories (64\% vs.\ 67\% mean performance) and identify bugs in the agent's implementation (17\% vs.\ 60\% success rate), while reducing perceived workload (e.g., frustration reduced from 5.4/7.0 to 2.4/7.0) compared to conventional tools.

\end{abstract}

\maketitle

\section{Introduction}

Large language models (LLMs) have demonstrated remarkable capabilities in generating~\cite{Chen2021,ziegler2022productivity}, editing code~\cite{Gupta2023,Bairi2024}, and testing code~\cite{Lemieux2023,Ryan2024,Yuan2024,Hayet2024a,icse2024-Fuzz4All}.
The most recent LLM-based software engineering techniques are agents that go beyond single-step prompting by autonomously interacting with their environment.
Such agents iteratively interact with their environment by generating LLM queries, interpreting the responses, invoking tools, and modifying code.
Software engineering agents are increasingly effective, e.g., in setting up programming environments~\cite{issta2025_ExecutionAgent,Milliken2025}, solving issues reported by users~\cite{Yang2024a,Zhang2024a}, and fixing bugs via program repair~\cite{icse2025-RepairAgent}.
For example, an agent may read an issue description, search for relevant code snippets, generate a patch, test the patch, and iterate until the issue is resolved.

While software development agents hold great promise for reducing manual effort, creating and improving the agents themselves remains challenging.
Developers of agents must design suitable prompts, implement the agent logic, and fix issues that arise during agent execution.
As with any piece of software, understanding and debugging software development agents is crucial for their correctness and effectiveness.
However, debugging such agents is particularly difficult due to the inherent non-determinism of LLMs and the complexity of an agent's interaction with the development environment.
In particular, developers are facing four key challenges when debugging software development agents:
\begin{itemize}
  \item \emph{C1: Prompt engineering.} Designing effective prompts for LLMs is a complex task that often requires trial and error. Developers need to understand how different prompt formulations affect the agent's behavior, and they want to receive quick feedback on prompt changes.
  \item \emph{C2: Understanding agent behavior.} Software development agents typically proceed in a loop of prompting an LLM, interpreting the response, invoking a tool suggested by the LLM, and providing the tool's output back to the LLM.
  Agents targeted at complex software development tasks often use specialized tools, e.g., for code search, code editing, and testing.
  Typical trajectories involve dozens of iterations, during which hundreds of thousands of tokens are exchanged with the LLM~\cite{ase2025_agent_study}, making it challenging to follow the agent's reasoning.
  \item \emph{C3: Resolving bugs in the agent program.} The \emph{agent program}, also called scaffold, orchestrates the interaction between the LLM and the development environment.
  Like any software system, an agent program may contain bugs that lead to incorrect or suboptimal behavior, which agent developers must identify and fix.
  \item \emph{C4: Reviewing intermediate code changes.} Software development agents often modify code in the target repository as part of their operation.
  To understand and debug the agent, developers need to review code changes made by the agent at different points during its executions, i.e., not only the final result.
\end{itemize}

The probably most common, state-of-the-practice way of addressing these challenges is to manually inspect the raw logs produced during agent execution.
Unfortunately, such logs are often unstructured and voluminous, making it difficult to extract relevant information.
Log viewers, e.g., offered by LLM platforms such as OpenAI~\cite{OpenAIPlatform} and LangChain~\cite{LangChain}, can help by providing search and filtering capabilities to navigate these logs more easily.
However, existing log viewers do neither address the unique challenges of software development agents, such as challenge C4, nor do they provide interactive debugging capabilities that allow developers to control the agent's execution.
Hence, the challenge of understanding and debugging software development agents remains largely unsolved.

This paper introduces \emph{\name{}}, the first interactive debugger specifically designed for LLM-based software development agents.
\name{} consists of three main components -- a user interface, a backend, and an API -- that work together to enable interactive debugging of agent trajectories.
First, the web-based user interface presents the trajectory of an agent as a structured conversation among the LLM, the agent program, and the tools invoked during execution.
Similar to conventional debuggers, \name{} supports breakpoints, stepwise execution, and live editing of prompts and tool invocations.
Additionally, the interface displays repository-level code changes at different points in time as a commit history, enabling developers to track how the agent modifies the code base.
Second, the debugger backend captures and stores events during an agent's execution, records intermediate code changes performed by the agent, and manages the resulting agent trajectories.
Third, the API enables attaching \name{} to existing agents with minimal code changes. By invoking the API at critical points in the agent program, such as before and after an LLM query or a tool call, developers can instrument their agents to interact with \name{}, similar to setting breakpoints in a conventional debugger.

The approach exploits our observation that debugging software development agents shares many similarities with conventional debugging of software programs.
Both activities involve understanding the flow of execution, inspecting intermediate states at key points during execution, and manipulating the execution to test hypotheses about a program's behavior.
While a conventional debugger can also be applied to the agent program itself, this would expose the developer to low-level implementation details that are often not relevant for understanding the agent's overall behavior.
Instead, \name{} raises the level of abstraction from the agent program's source code to the high-level actions performed by the agent when interacting with the LLM and tools.
By adopting established concepts of interactive debugging to the domain of software development agents, \name{} provides developers with a powerful tool to understand and control the behavior of these agents.
In practice, developers may, of course, combine \name{} with conventional debuggers to inspect and debug the agent program itself when needed, which we leave for future work to explore.

We evaluate \name{} by integrating it into three state-of-the-art software development agents: ExecutionAgent~\cite{issta2025_ExecutionAgent}, SWE-Agent~\cite{Yang2024a}, and RepairAgent~\cite{icse2025-RepairAgent}.
The integration requires only minor modifications to the agent programs, with 5--7 API calls inserted and 39--42 lines of code changed, demonstrating the ease of adoption.
To assess the effectiveness of \name{} in supporting developers, we conduct a user study with twelve participants.
Given a trajectory comprehension task and two agent debugging tasks, the participants either use \name{} or inspect a raw log with a tool of their choice.
We observe that participants using \name{} are better able to understand the agent's behavior (with the mean performance increasing from 64\% to 67\%), and that they identify bugs in the agent program more effectively (with the success rate increasing from 17\% to 60\%).
Moreover, participants using \name{} report a lower perceived workload compared to those participants using conventional tools, e.g., with perceived frustration reduced from 5.4/7.0 to 2.4/7.0.

In summary, this paper contributes the following:
\begin{itemize}
  \item We identify the unique challenges developers face when understanding and debugging software development agents.
  \item We draw parallels between debugging conventional software and debugging software development agents, laying the foundation for our approach.
  \item We present the first interactive debugger for LLM-based software development agents, which enables developers to inspect, control, and manipulate agent trajectories.
  \item We evaluate our approach by integrating it into three state-of-the-art software development agents and by conducting a user study, demonstrating its effectiveness in supporting developers.
  \item We make our code and data publicly available to foster further research in this area.
\end{itemize}

\section{Background on Software Development Agents}
\label{sec:background}

The term ``agent'' is used in different ways in the literature, sometimes simply meaning any program that uses an LLM.
Following prior work~\cite{icse2025-RepairAgent}, we define an LLM-based software development agent as a program with two properties:
First, the agent relies on an LLM to autonomously plan and execute a sequence of actions to achieve a goal.
Notable, this ``agency'', i.e., the ability to make choices and act on them, distinguishes agents from techniques that rely on hard-coded LLM prompts and techniques that follow a hard-coded algorithm to issue a pre-defined sequence of LLM queries.
Second, the agent interacts with its environment by performing actions suggested by the LLM.
These actions typically are invocations of tools that enable the agent to interact with a development environment similar to a human developer.

Based on this definition, an agent can be seen as having three components: (i) the core program that orchestrates the interaction between the LLM and the environment, which we call the \emph{agent program},\footnote{Other terms used in the literature include ``scaffold''~\cite{DBLP:conf/icml/Pan0NJ0S025}, ``agent framework''~\cite{yin2025comprehensiveempiricalevaluationagent}, ``middleware''~\cite{icse2025-RepairAgent}, and ``orchestrator''~\cite{chen2025agentguard}.} (ii) the \emph{LLM} that is prompted by the agent program to suggest the next action, and (iii) the \emph{tools} invoked by the agent program, as suggested by the LLM.
Recent work has proposed various software engineering agents that follow this paradigm, targeting different tasks, such as setting up programming environments~\cite{issta2025_ExecutionAgent,Milliken2025,Eliseeva2025,Yang2025,hu2025llm}, solving issues reported by users~\cite{Zhang2024a,Wang2024a,Yang2024a,gao2025trae}, and fixing bugs via program repair~\cite{icse2025-RepairAgent}.
The most effective agents are typically equipped with specialized tools tailored to the specific task, such as different kinds of code search tools, code editing tools, tools for running tests, and tools for applying static analyses.

\section{Approach}

The following presents \name{}, our approach for interactively debugging software development agents.
The conceptual underpinning of the approach is to adapt established concepts from conventional debugging to the domain of software development agents.
To this end, Section~\ref{sec:parallels_conventional_debugging} discusses similarities and differences between debugging conventional software and debugging software development agents.
Next, Section~\ref{sec:approach_overview} provides an overview of the approach, followed by detailed descriptions of the three main components: the user interface (Section~\ref{sec:user_interface}), the backend (Section~\ref{sec:backend}), and the API for integrating \name{} into existing agents (Section~\ref{sec:api}).

\subsection{Similarities and Differences Compared with Conventional Debugging}
\label{sec:parallels_conventional_debugging}

As LLM-based agents are programs, there are, in principle, amenable to traditional approaches for understanding and debugging programs.
However, because agents make heavy use of LLMs, large parts of an agent's behavior are not explicitly programmed, but emerge from the interaction of the agent program with the LLM and the environment.
That is, attaching a traditional debugger to the agent program exposes low-level implementation details, e.g., how exactly to invoke a remotely hosted LLM, but it reveals little about the high-level reasoning of the agent as it interacts with the LLM and tools.
At the same time, traditional debugging tools offer valuable concepts, such as breakpoints, stepwise execution, and live editing of program state, which could also benefit developers of software development agents.

To benefit from these concepts when debugging software development agents, we identify two parallels between developing conventional software and developing software development agents:
\begin{itemize}
  \item \emph{Choosing and implementing algorithms} $\approx$ \emph{Designing effective prompts.} Developers must choose appropriate algorithms and implementation strategies when developing traditional software.
  When developing a software development agent, large parts of this activity are replaced by designing effective prompts, as these prompts effectively guide the LLM toward solving the task at hand.
  For example, a prompt that instructs the LLM to first search for existing code before generating new code will likely lead to different behavior than a prompt that directly asks the LLM to generate code.
  
  \item \emph{Understanding control flow and data flow} $\approx$ \emph{Understanding agent behavior.}
  In conventional software development, developers need to understand the control flow and data flow in their program to reason about its behavior.
  Similarly, developers of software development agents must understand how the agent interacts with the LLM and tools over time to reason about the agent's behavior.
  For example, understanding which tools an agent invokes at which point in time, and how the LLM reacts to outputs produced by these tools is crucial for comprehending and ultimately improving the agent's reasoning.
\end{itemize}

Based on these parallels, the key hypothesis of our work is that adapting established concepts from conventional debugging to software development agents can improve the ability of agent developers to understand and debug these agents.
Importantly, instead of merely copying these concepts, we need to adapt them to the unique characteristics of software development agents.
This adaptation requires raising the level of abstraction from the agent program's source code to the high-level actions performed by the agent, which we address as presented in the following.

\subsection{Overview of Approach}
\label{sec:approach_overview}

\begin{figure}
  \includegraphics[width=1\linewidth]{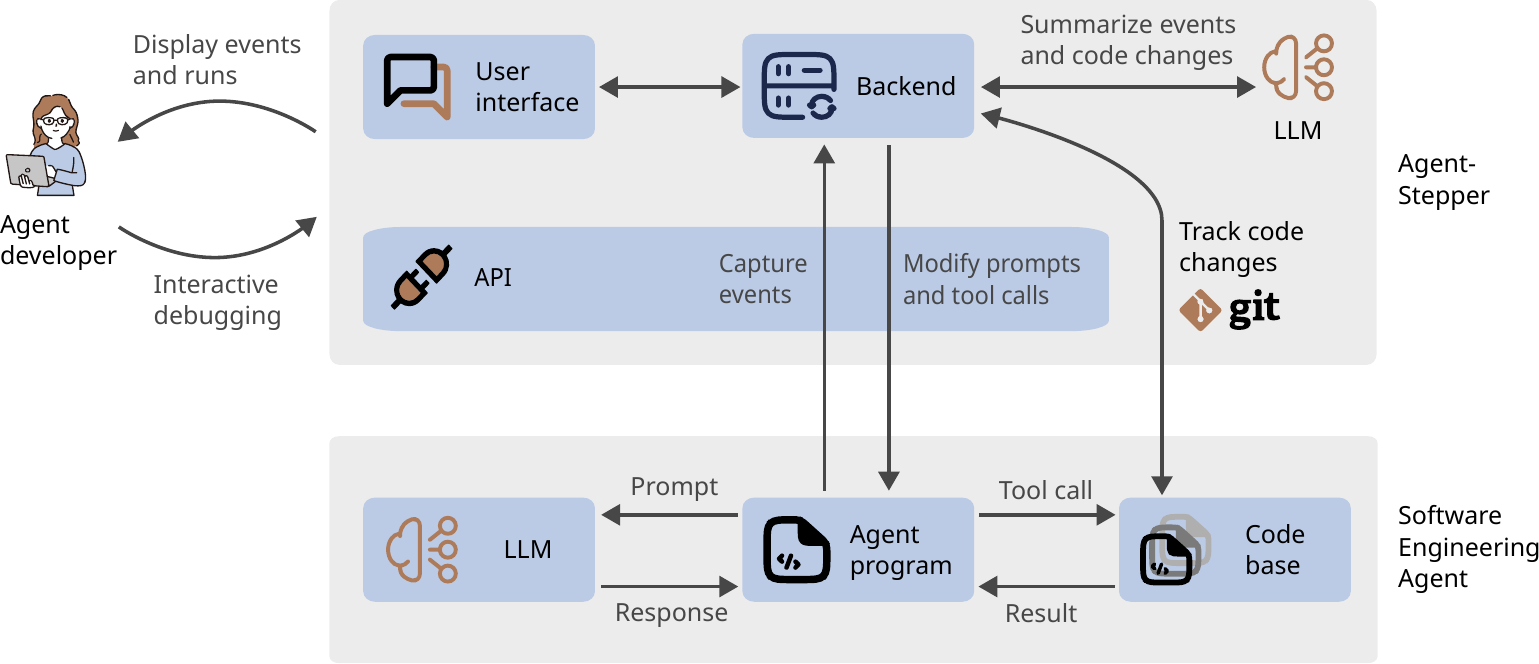}
  \caption{Overview of the approach. The upper part of the figure (\name{}) is this paper's contribution.}
  \label{fig:overview}
\end{figure}

Figure~\ref{fig:overview} presents an overview of our approach.
The upper part of the figure shows \name{}, i.e., our novel approach for interactively debugging software development agents.
Such agents, as shown in the lower part of the figure and described in Section~\ref{sec:background}, consist of an agent program that orchestrates the interaction between an LLM and tools applied to the code base.
To understand and debug such agents, an agent developer uses \name{}, which consists of three main components: a user interface, a backend, and an API for integrating \name{} into existing agents.
To present the voluminous information produced during an agent's execution, such as lengthy prompts and potentially large code changes, in a concise manner, \name{} itself also uses an LLM to summarize that information.
Our approach also builds upon a version control system, such as git, to track intermediate code changes made by the agent during its execution.

\subsection{User Interface}
\label{sec:user_interface}

The user interface presents agent trajectories in a clear format through four key design choices: (1) a conversation-based representation that structures agent-LLM and agent-tool interactions as interleaved conversations (Section~\ref{sec:ui_conversations}), (2) balancing high-level overview with detailed inspection (Section~\ref{sec:summary_details}), (3) viewing code changes at different points in time (Section~\ref{sec:ui_code_changes}), and (4) interactive debugging via breakpoints, stepping, and editing (Section~\ref{sec:ui_interactive}).
The interface also supports managing and importing past agent runs.

\begin{figure}
  \includegraphics[width=\linewidth]{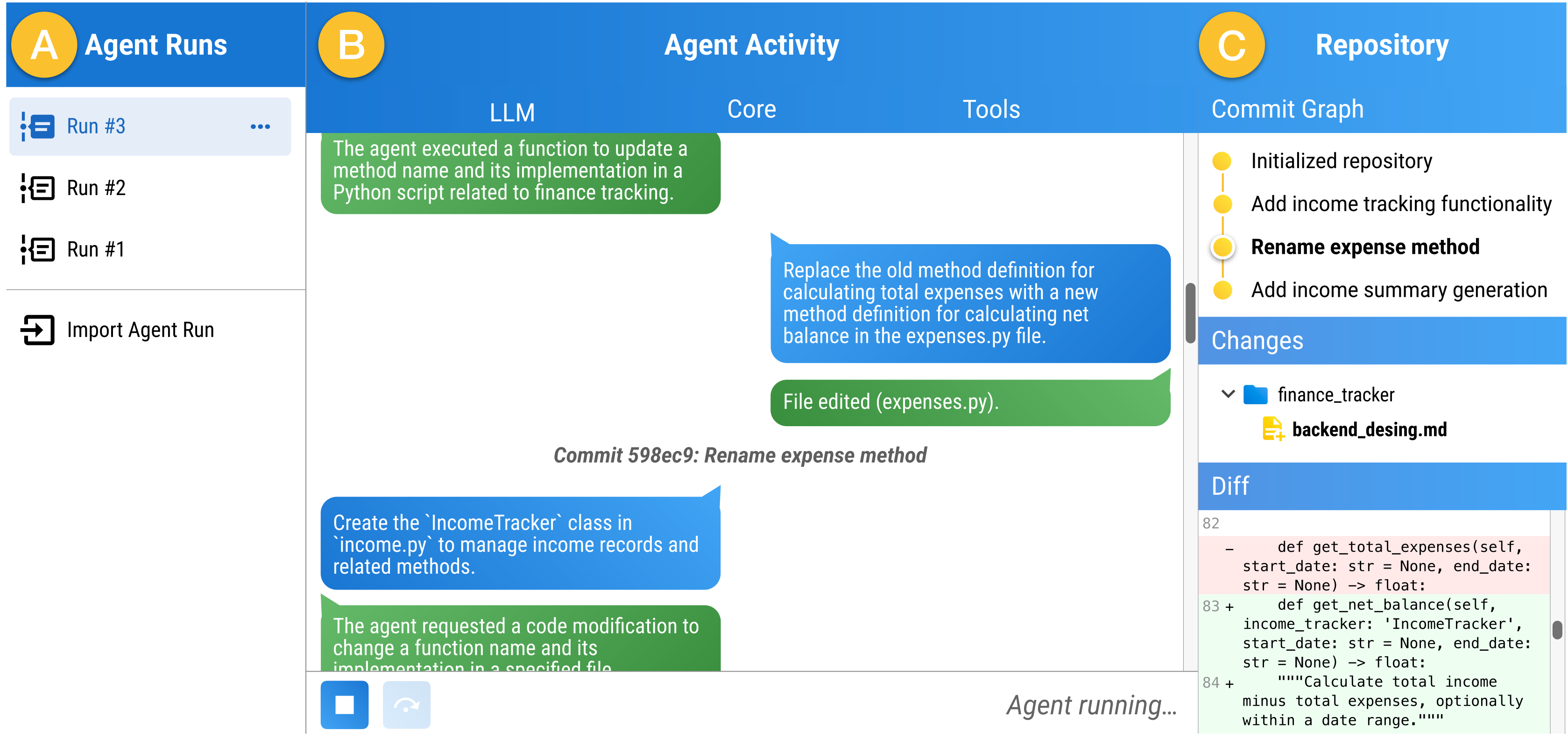}
  \caption{User interface of \name{}. Part~A is a panel to select agent runs. Part~B shows the structured conversation view with breakpoints and stepping controls. Part~C displays repository-level code changes.}
  \label{fig:screenshot}
\end{figure}

Figure~\ref{fig:screenshot} shows a screenshot of the user interface of \name{}.
Part~A presents a panel for selecting different runs of an agent.
Part~B displays the structured conversation view, showing the interleaved conversations between the agent program and the LLM, and between the agent program and the tools.
At the bottom of part~B, the agent control panel allows developers to stop at breakpoints, step through the execution, and continue execution.
Part~C shows the repository-level code changes made by the agent during its execution, presented as a commit history.

\subsubsection{Structured Conversations}
\label{sec:ui_conversations}

Agent execution produces raw logs that present a single long sequence of prompts, responses, tool calls, and tool outputs, which poses challenges for developers trying to understand agent behavior.
While raw logs contain all information about an agent's execution, they present many low-level details that make it difficult to follow the agent's reasoning.
For example, a recent study reports a mean token count per agent run ranging from 23k to 1.2M, depending on the agent, and a typical run involves several dozens of cycles that each invoke a tool producing some output~\cite{ase2025_agent_study}.
Log viewers, such as those provided by OpenAI~\cite{OpenAIPlatform} and LangChain~\cite{LangChain}, improve upon raw logs by offering search and filtering capabilities, but they still present information as a linear sequence of events without higher-level structure.

To address these limitations, \name{} introduces a structured representation of agent trajectories that organizes agent interactions into two interleaved conversations: one between the agent program and the LLM, and another between the agent program and the tools.
This representation follows the metaphor of a chat application, similar to interfaces commonly used in modern chatbot systems, but extends this metaphor to simultaneously display two conversations that proceed in parallel.
The two conversations are naturally interleaved, as the agent program performs its typical four-step cycle:
(i) the agent program sends a prompt to the LLM, and (ii) the LLM responds;
these first two steps are part of the conversation between the agent program and the LLM.
Then, (iii) the agent program invokes a tool based on the LLM's response, and (iv) the tool returns its output to the agent program;
these last two steps are part of the conversation between the agent program and the tools.
By organizing messages around these cycles, we provide a natural and intuitive structure for presenting the interleaved conversations.
Messages in both conversations are displayed side-by-side in two columns and sorted chronologically, with the most recent messages appearing at the bottom, enabling developers to follow the agent's interactions over time.
The screenshot in Figure~\ref{fig:screenshot}, part B, illustrates this structured conversation view.

\subsubsection{Summarized and Detailed Views of Events}
\label{sec:summary_details}

An important challenge is to present the voluminous information produced during an agent's execution in a concise manner that still conveys the essential details.
This challenge stems from the fact that prompts, responses, tool call arguments, and results are typically lengthy and often presented in raw formats (e.g., JSON) that are unsuitable for direct display as readable text.
\name{} addresses this challenge by summarizing messages to provide agent developers with a comprehensive overview of an agent's activities throughout its execution.
The summaries are displayed in the chat bubbles of the conversation view, as shown in Figure~\ref{fig:screenshot}, part B.

To summarize the individual events that occur during an agents 4-step cycle, \name{} prompts an LLM to condense the details of an event into a single sentence.
In the summarization prompt, we instruct the LLM to focus on event-specific content while disregarding boilerplate or repetitive information that does not contribute to understanding the agent's behavior.
We use four different summarization prompts, each tailored for one of the four message types (LLM queries, LLM responses, tool invocations, and tool outputs), which are provided in the supplementary material.
As prompts issued by agents often include standard instructions or repeated sections, it is important that the summaries focus on the unique and relevant aspects of each message.
To achieve this goal, we provide the LLM with both the current event and the preceding event, prompting it to highlight only the differences between the two events in the summary.

While the summaries provide a high-level overview of the agent's activities, developers must sometimes inspect the full details of specific messages to understand how the agent program invokes tools or interprets responses.
To support this need, \name{} provides a message inspector window that displays comprehensive message details.
The inspector window is structured similarly to a typical email client, presenting both meta-information about the message, including its type, origin, and destination, as well as the full content of the message.
Message content is rendered in plain text format for simple messages, or through an interactive JSON object viewer for messages represented as JSON or dictionaries, enabling developers to navigate and understand complex structured data.
In addition, the message inspector provides a comparison feature that displays the differences between the selected message and another message, which is particularly useful when comparing subsequent prompts where specific sections are updated dynamically based on the previous cycle.

\subsubsection{Repository-Level Code Changes}
\label{sec:ui_code_changes}

Agents employed in software development primarily operate on code bases, where the agent may modify existing source files, create new ones, remove outdated files, build a project, and execute tests. 
When operating autonomously, agents often introduce numerous changes to the code base over the course of their trajectory. 
However, these intermediate modifications remain invisible to the user, as agents typically work on a local copy of the code and transfer changes only if they consider the run successful. 
Consequently, the user receives either the final submission or nothing in the case of a failed run. 
While this all-or-nothing approach may be suitable for end users, it poses significant challenges for agent developers aiming to understand and debug the agent's behavior.

\name{} addresses the challenge of reviewing intermediate code changes by visualizing repository-level modifications made by the agent during its execution in a dedicated side panel (part~C in Figure~\ref{fig:screenshot}).
This presentation is inspired by commit history views offered by version control systems and IDEs. 
Each modification made by a tool invocation is represented as a commit in the history, with a commit message summarizing the change. 
Developers can click on a commit to view the exact changes made to the code base in a diff viewer. 
In addition, links to the corresponding commits are provided in the main conversation view (see center part of Figure~\ref{fig:screenshot}), allowing developers to quickly navigate to the relevant code changes associated with cycles of the agent's execution.
Importantly, the commit history created by \name{} is independent of the version control system used by the target repository: While the latter is typically updated once an agent has completed its run, the former is updated after each tool invocation, providing a fine-grained view of the agent's modifications throughout its execution.

\subsubsection{Breakpoints, Stepping, and Live Editing of Prompts and State}
\label{sec:ui_interactive}

\name{} can be used in two modes: either \emph{post hoc} to inspect a completed trajectory or \emph{interactively} to control the execution of an agent in real time. 
The following presents the interactive mode.
Following the analogy with conventional debugging, a key requirement of an interactive debugger is the ability to halt the program at breakpoints and step through execution event by event. 
In the user interface, this functionality is facilitated by a control panel, as shown in the lower end of part B in Figure~\ref{fig:screenshot}.

\name{} provides a flexible breakpoint system that allows developers to control the execution of events during debugging. 
By default, each event defines two breakpoints: before and after the event is executed. 
Developers can adjust this default behavior using the API described in Section~\ref{sec:api}. 
When the execution reaches a breakpoint, \name{} pauses the agent's execution and displays the corresponding event in the conversation view. 
Similar to traditional debuggers, developers can either step through the execution one breakpoint at a time or continue execution until they decide to return to stepping mode.

\name{} not only allows developers to observe the agent's execution but also enables them to interactively modify the agent's behavior at breakpoints. 
This functionality is akin to modifying variable values in a traditional debugger.
Such interactive capabilities are useful in several scenarios. 
First, developers can engage in interactive prompt engineering by modifying a prompt before it is sent to the LLM, thereby testing how the change affects the agent's behavior. 
Second, they can simulate LLM responses by editing the actual response from the LLM to evaluate how the agent program handles different responses. 
Third, the ability to modify tool invocations allows developers to change the arguments of a tool call or replace the tool call suggested by the LLM with a different tool call. 
Finally, developers can simulate tool outputs by editing the output returned by a tool, which facilitates testing how the agent program responds to various outputs.

\begin{table}
  \caption{API functions to integrate \name{} into existing software development agents.}
  \label{tab:api}
  \small
  \begin{tabular}{@{}lp{15em}p{25.5em}@{}}
    \toprule
    Id & API function & Description \\
    \midrule
    1 & \scode{\_\_init\_\_(agent\_name: str, address: str, port: int, workspace\_path: Optional[str])} & Establishes a connection to the backend and initializes a new \scode{AgentStepper} object. The \scode{agent\_name} parameter identifies the agent as a whole, while \scode{address} and \scode{port} specify the backend's location. Optionally, a workspace path can be provided to track changes if the agent edits a local project directory. \\
    \midrule
    2 & \scode{begin\_llm\_query\_breakpoint(prompt: Union[str, Dict]) -> Union[str, Dict]} & Marks the beginning of an LLM query event. The method sends the query prompt to the debugger and halts execution if a breakpoint is triggered. The user may modify the prompt in the UI, in which case the updated prompt is returned. If no modifications are made, the original prompt is returned unchanged. \\
    \midrule
    3 & \scode{end\_llm\_query\_breakpoint(response: Union[str, Dict]) -> Union[str, Dict]} & Marks the completion of an LLM query event. The method transmits the response received from the LLM to the debugger and halts execution if required. As with the prompt, the user may modify the response via the UI. The method then returns either the modified or the original response. \\
    \midrule
    4 & \scode{begin\_tool\_invocation\_breakpoint( tool: str, args: Dict) -> Tuple[str, Dict]} & Encapsulates the beginning of a tool invocation event. The tool identifier and its arguments are sent to the debugger, allowing the user to inspect and possibly alter them before execution. The return value consists of the (potentially updated) tool identifier and argument dictionary. \\
    \midrule
    5 & \scode{end\_tool\_invocation\_breakpoint( results: Union[str, Dict]) -> Union[str, Dict]} & Marks the end of a tool invocation event. The results produced by the tool are provided to the debugger and may be altered in the user interface before being returned to the agent. \\
    \midrule
    6 & \scode{commit\_agent\_changes( commit\_summary: str, commit\_description: str) -> bool} & Commits the current state of the agent's workspace to its version control repository and synchronizes the commit with the \name{} backend. If no summary or description is supplied, they are generated automatically. The method returns \scode{True} if changes were committed, and \scode{False} otherwise. \\
    \midrule
    7 & \scode{post\_debug\_message(message: str)} & Sends a textual message to the debugger user interface. This method can be used by the agent program to provide additional contextual information or debugging notes to the developer. \\
    \bottomrule
  \end{tabular}
\end{table}

\subsection{Backend}
\label{sec:backend}

The backend of \name{} captures and manages agent events during execution, records intermediate code changes, and manages agent trajectories. 
Events are triggered via API calls performed by the agent program, e.g., whenever the agent program sends a prompt to the LLM. 
To record intermediate states, the backend initializes the agent's workspace as a git repository at the beginning of the run and creates a new branch for the run. 
After the run concludes, the API automatically switches back to the original branch in preparation for the next execution. 
After each tool invocation, the backend commits the pending changes to the repository. 
To obtain a commit message, the backend queries an LLM with the diff produced by the tool invocation and asks it to generate a concise message describing the change.

In interactive mode, \name{} manages the execution state of the agent by supporting three distinct states to control its progress. 
The first state is ``paused,'' where the agent is halted at a breakpoint, allowing developers to inspect the current state. 
The second state is ``stepping,'' in which the agent executes one event at a time and waits for user input after each event, facilitating a detailed examination of the agent's behavior. 
Finally, the ``running'' state enables the agent to execute continuously until the developer decides to pause or step through the execution.

\subsection{API for Integrating into Agents}
\label{sec:api}

To capture events during an agent's execution and enable interactive debugging, \name{} provides an API that developers of agents can use to instrument their agent programs.
By allowing developers to insert API calls at critical points in the agent program, the approach raises the level of abstraction from the agent program's source code to the high-level actions performed by the agent when interacting with the LLM and tools.
We evaluate the ease of integrating \name{} into existing agents in Section~\ref{sec:evaluation}, showing that only a few dozens of code lines need to be changed to add support for \name{}.

Table~\ref{tab:api} presents the seven API functions that \name{} provides to integrate with existing software development agents.
The API is offered by a single \code{AgentStepper} class that developers can instantiate in their agent program to establish a connection to the \name{} backend (entry~1 in the table).
The next four entries (2--5) correspond to breakpoints that developers can insert before and after LLM queries and tool invocations.
These four API functions enable \name{} to capture the agent's interaction with the LLM and tools, and they support interactive editing of prompts, responses, tool calls, and tool outputs.
Entry~6 in the table allows the agent program to commit changes made to the agent's workspace to the version control repository managed by \name{}.
By default, \name{} automatically commits changes after each tool invocation, but developers can also use this API function to commit changes at other points in time, e.g., in case the agent program modifies the code base without invoking a tool.
Finally, entry~7 enables the agent program to post arbitrary debug messages to the \name{} user interface, which developers can use to provide additional context or debugging notes.

\begin{figure}
  \begin{lstlisting}[numbers=left]
class MyAgent(Agent):
  def think(self) -> Action:
    prompt = self.get_next_prompt()(@\label{line:get_prompt}@)
    (@\HL@)prompt = self.debugger.begin_llm_query_breakpoint(prompt)(@\HLoff@)
    response = self.llm.get_completion(prompt)(@\label{line:llm_query}@)
    (@\HL@)response = self.debugger.end_llm_query_breakpoint(response)(@\HLoff@)
    return self.response_to_action(response)
		
def main():
  agent = MyAgent()
  with (@\HL@)AgentStepper('MyAgent', 'localhost', 8765, 'agent_workspace')(@\HLoff@) as debugger:(@\label{line:agent_stepper}@)
    agent.debugger = debugger
    while not agent.is_done():(@\label{line:main_loop}@)
      action = agent.think()
      (@\HL@)(action.name, action.args) = debugger.begin_tool_invocation_breakpoint(action.name, action.args)(@\HLoff@)
      result = environment.execute(action)(@\label{line:action}@)
      (@\HL@)result = debugger.end_tool_invocation_breakpoint(result)(@\HLoff@)
      agent.add_observation_to_history(result)(@\label{line:main_loop_end}@)
  \end{lstlisting}
  \caption{Minimal agent program using the \name{} API. Code with gray background shows API calls.}
  \label{lst:api_example}
\end{figure}

To illustate how to use the API, consider the minimal agent program shown in Figure~\ref{lst:api_example}.
The agent consists of a main loop (lines~\ref{line:main_loop}--\ref{line:main_loop_end}) that repeatedly calls the \code{think} method to determine the next action to perform and then executes that action.
The \code{think} method first constructs the next prompt to send to the LLM (line~\ref{line:get_prompt}) and then invokes the LLM to obtain a response (line~\ref{line:llm_query}).
To use our approach, the agent program connects to \name{} by creating an \code{AgentStepper} object (line~\ref{line:agent_stepper}).
The agent program then invokes the API via this object: before and after each LLM query (highlighted code around line~\ref{line:llm_query}), and before and after each tool invocation (highlighted code around line~\ref{line:action}).
Instead of propagating all events to \name{}, as done in this simple example, agent developers may also expose events selectively, giving them finer control over which events to inspect and manipulate during debugging.

\section{Evaluation}
\label{sec:evaluation}

We evaluate our work by applying \name{} to three state-of-the-art software development agents and by conducting a user study to assess its effectiveness in supporting developers.
The evaluation addresses the following research questions:

\begin{itemize}
  \item RQ1: What effort is required to integrate \name{} into existing software development agents?
  \item RQ2: To what extent does \name{} support developers in understanding and debugging software development agents?
  \item RQ3: How does using \name{} affect the perceived workload of developers?
\end{itemize}

\subsection{Experimental Setup}

\paragraph{Implementation}
We implement our approach to be compatible with state-of-the-art LLM agents and frameworks, such as SWE-Agent~\cite{Yang2024a}, OpenHands~\cite{Wang2024a}, RepairAgent~\cite{icse2025-RepairAgent}, and Execution\-Agent~\cite{issta2025_ExecutionAgent}.
Since these projects are developed in Python, we implement both the API and the backend of \name{} in Python.
The user interface is a web application implemented in Vue.js and is hosted on a web server started by the backend.
For implementing the communication between the API and the backend, as well as the backend and the user interface, we use WebSockets, as they support bidirectional messaging and maintain a persistent connection throughout the agent's execution.

\paragraph{Agents}
We integrate \name{} into three state-of-the-art, open-source software development agents that address three different tasks and are implemented in different ways.
First, we use SWE-Agent~\cite{Yang2024a}, which is multi-purpose but was originally intended to help developers resolve problem reports by automatically generating code patches.
As of September 2025, it was ranked as the top-performing agent on SWE-bench~\cite{Jimenez2024}.
Second, we use RepairAgent~\cite{icse2025-RepairAgent}, which focuses on automated program repair by generating patches for buggy code based on failing test cases.
It takes as an input a code base with failing tests and iteratively suggests code changes until all tests pass or a maximum number of iterations is reached.
As of its publication, it was the state-of-the-art agent for program repair on the Defects4J benchmark~\cite{Just2014}.
Third, we use ExecutionAgent~\cite{issta2025_ExecutionAgent}, which is designed to autonomously set up programming environments by installing dependencies and configuring settings.
Its input is the URL of a GitHub repository, and it aims to create a containerized environment where the test suite of the repository can be executed successfully.
As of its publication, ExecutionAgent was the first agent to address this task, which has also been addressed by concurrent and follow-up work~\cite{Milliken2025,Eliseeva2025,Yang2025,hu2025llm}.
These agents differ in their tasks, architecture, and the tools they invoke, providing a diverse set of case studies for evaluating \name{}.

\subsection{RQ1: Integration Effort}

To integrate \name{} into each of the three agents, we familiarize ourselves with their implementations (a step that is not necessary for developers of an agent, who are the prime audience for a debugging technique like \name{}) and then insert API calls at specific locations in the agent program.
Specifically, we instantiate the \code{AgentStepper} object at a point before entering each agent's main execution loop, establishing the connection to the \name{} backend before any agent actions are performed.
We then insert API calls at critical points in the agent program: before and after each LLM query, before and after each tool invocation, and at locations where the agent modifies the code base.
To ensure seamless integration with the user interface, we implement additional code that converts data structures internal to the agent program, such as agent-specific prompt representations and tool result formats, into formats compatible with the \name{} API.
This conversion ensures that data produced by the agent is displayed correctly in the user interface and that changes applied by developers in the user interface are consistently propagated back to the agent program during interactive debugging sessions.

To quantitatively assess the integration effort, we employ three metrics: the number of API calls added, the number of files modified, and the lines of code changed (added, deleted, or modified).
These metrics provide an objective and quantifiable measure of the work required to integrate \name{} into an agent program.
We select these metrics over alternative approaches, such as measuring integration time, which would be more subjective and influenced by factors including developer familiarity with the agent code base and the complexity of the agent's architecture.

\begin{table}[t]
\centering
\caption{Overview of code changes made to integrate \name{} into agents.}
\label{tab:integration_effort}
\small
\begin{tabular}{@{}lrrr@{}}
\toprule
Agent & API calls & Files modified & Lines changed\\
\midrule
SWE-Agent       & 7 & 3 & 42 \\
RepairAgent     & 7 & 5  & 39 \\
ExecutionAgent   & 5  & 4  & 39 \\
\bottomrule
\end{tabular}
\end{table}
Table~\ref{tab:integration_effort} summarizes the results of integrating \name{} into the three agents. 
Each agent requires between 5 to 7 API calls to be added.
The reason for these relatively small numbers is that agent programs typically dispatch LLM queries and tool invocations through centralized functions or methods, allowing us to insert API calls in just a few locations to capture all relevant events.
The modifications performed to integrate \name{} into the agents affects between 3 and 5 code files per agent, where the total lines of code changed per agent ranges between 39 and 42. 
Overall, these findings indicate that the integration process is manageable and does not impose a significant burden on developers.

\subsubsection{Case Study 1: SWE-Agent}


To offer a more qualitative perspective on the integration effort, we present two case studies.
The first case study focuses on SWE-Agent, where we made a total of 42 lines of code changes across 3 files.
To instrument SWE-Agent, we first insert API calls for LLM queries and tool invocations. The \code{AgentStepper} object is instantiated in \code{run\_single.py} before the agent starts, ensuring it persists throughout the agent's execution. For debugging LLM queries, API calls are added in the \code{query} method of \code{models.py} before and after fetching the LLM completion. The entire message history, which consists of prior queries and tool invocations, is passed as a JSON-serializable dictionary to the \code{begin breakpoint} API call. This format enhances readability in the user interface. If modifications are made, they are directly written to the internal \code{messages} variable. Ending the query breakpoint involves passing the completion result as a JSON structure to the debugger, allowing for either the original or modified result to be returned.

Handling tool calls and their effects is more complex because SWE-Agent runs within an isolated Docker container.
The container starts at the beginning of execution, and the agent copies the user's code repository into it.
Throughout execution, the agent executes commands within this isolated environment, modifying only the container's copy of the code.
Normally, SWE-Agent transfers the changes to the user's file system and applies them to the original repository only if the agent successfully generates a submission.
We enable \name{} to capture tool invocations and results despite this isolation by modifying the \code{forward} method in \code{agents.py}.
The modified code captures the tool invocation before it is being sent to the container and sends it to the debugger via the \code{begin\_tool\_invocation\_breakpoint} API call as a a JSON-serializable dictionary.
After tool execution, we capture the result and send it to the debugger to end the breakpoint.
To track code modifications made within the Docker container, we utilize SWE-Agent's submission mechanism to generate and apply a patch to the local repository after each tool invocation.

\subsubsection{Case Study 2: RepairAgent}


The second case study focuses on RepairAgent, where we made a total of 39 lines of code changes across 5 files.
RepairAgent is built on top of the AutoGPT framework and operates in an isolated devcontainer.
To integrate \name{} into RepairAgent, the first task is to modify the devcontainer's configuration file. 
This allows the agent program to connect to \name{} by adding a line to \code{devcontainer.json} that instructs Docker to use the host machine's network. 
By sharing the host machine's network, the agent program can communicate directly with \name{}'s backend.

We instantiate the \code{AgentStepper} object in the \code{run\_interaction\_loop} function of\linebreak\code{main.py}, which implements the agents cyclic operation.
To set a breakpoint at each tool invocation, we enclose the \code{agent.execute()} statement in API calls within the loop of the \code{run\_interaction\_loop} function.
Because the command name is stored as a string and the command arguments are stored as a JSON-serializable dictionary object, we pass these values to the API directly without conversion.
Consequently, we can apply modifications directly.
The same applies to the result of the tool invocation, which is stored as a string.
In contrast, the prompts are stored in an object that is not a dictionary, thus we convert the object into a JSON-serializable dictionary before passing it to the \code{begin breakpoint} call.
To apply modifications, we convert the modified dictionary back into a \code{ChatSequence} object.
Since the LLM completion is returned as a string, we pass it directly to the \code{end breakpoint} call, and we apply modifications directly.
To record the individual attempts made at fixing the bug by RepairAgent, an additional step is necessary.
Specifically, we add a call to \code{commit\_agent\_changes} right after the proposed fix is written to the code base but before RepairAgent may revert it in case the fix fails the test cases.

\subsection{RQ2: Usefulness for Understanding and Debugging Agents}

The primary objective of \name{} is to support developers in understanding and debugging software development agents.
To evaluate the effectiveness of \name{} in achieving this objective, we conduct a user study examining to what extent our approach supports users in comprehending agent trajectories and in locating bugs.
The study involves twelve participants, three tasks, and three software developments agents.
It is a between-group study, in which there are two groups of participants: the \emph{debugger group} that completes tasks using \name{} and the \emph{control group} that completes tasks using conventional methods.

\subsubsection{Methodology}

\paragraph{Tasks}

We design three tasks that mimic two typical kinds of problems encountered during the development of software development agents.
The first kind of problem, called \emph{trajectory comprehension}, involves understanding and interpreting the behavior of an agent based on its trajectory.
Specifically, participants analyze a partially successful agent trajectory, interpret it, and then answer a series of questions about the agent's behavior.
The second kind of problem, called \emph{bug identification}, involves locating and diagnosing bugs in an agent program based on a trajectory of the agent.
Participants are given a trajectory where an agent fails to achieve its goal due to a bug in the agent program's logic, and are asked to identify and describe the bug.
We design one trajectory comprehension task and two bug identification tasks, each involving a different software development agent.

For the trajectory comprehension task, participants review a trajectory of \emph{SWE-Agent}.
In this scenario, the agent receives a feature request to add income-tracking functionality to a personal finance tracker CLI application in Python, along with specific requirements, including a description of a new class and modifications to existing sources.
During execution, the agent successfully adds the new sources but becomes stuck in a loop when editing an existing file, repeatedly applying the same modification until termination.
To assess participants' understanding of the agent's behavior, they are asked to answer 14 questions about the trajectory.
We include both high-level questions concerning the entire trajectory, such as the agent's goals, whether it completed its task, and the key challenges encountered, and low-level questions focusing on specific cycles. 
Many of these questions address key trajectory characteristics, including recurring action sequences and the semantic coherence linking thoughts and actions across multiple cycles~\cite{ase2025_agent_study}.
For example, questions posed in this regard include: \textit{``Analyze the final four changes the agent makes to the repository. How do these changes relate to each other?"} or \textit{``What feedback does the agent receive after its first attempt to run the test script for the finance tracker, and how does it influence the next action?"}
Furthermore, we ask for precise details regarding individual changes, such as \textit{``What specific changes does the agent make to the cli.py file when it first modifies it?"} and \textit{``Which Python packages does the agent install over the course of the run?"}
Participants are given 25 minutes to answer these questions, either in free-form or by copying relevant information from the logs. 

For the first of the two bug identification tasks, participants analyze a trajectory from \emph{RepairAgent}.
The agent attempts to fix an issue in the Apache CSV Java library, but the agent fails due to a bug in the agent program that prevents it from applying its suggested changes to the code base.
The second bug identification task involves a trajectory from \emph{ExecutionAgent}.
In this scenario, the agent attempts to build the Google Gson project and run its test suite, but the agent fails because it cannot connect to a running Docker container due to a bug in the agent program.
Participants are given 12 minutes to identify and describe the bug that prevents the agent from succeeding.

\paragraph{Task Evaluation}

As all answers are given in free-form, we develop a standardized grading scheme and then use it to evaluate responses.
For the trajectory comprehension task, we grade each of the 14 questions individually.
For each question, we create an expected answer and determine a total number of points that can be awarded for an answer.
For example, a binary question gives either zero or one point, whereas a more complex question may give up to five points depending on the level of detail provided in the answer.
The grading scheme is initially developed by the first author of this paper, and then reviewed and refined together with another author to reduce subjectiveness and bias.
After grading, we normalize points by assigning each question the same weight to ensure all questions contribute equally to the total score.
For the bug identification tasks, there is only a single question, graded as correct or incorrect depending on whether the  participant correctly identifies the bug in the agent program.

\paragraph{Participants and groups}

We recruit twelve participants for the user study: half are active computer science students enrolled in bachelor's or master's programs, and half are PhD candidates in computer science.
Upon registering for the study, all participants complete a form detailing their demographics and prior technical experience in software development and LLM agent development.
Based on the information about the technical background of participants, we assign each participant to either the control group or the debugger group.
Participants in the debugger group complete the tasks using \name{}, where the trajectory is loaded into the debugger interface.
The participants are asked to use only the features of \name{} to explore the agent's behavior.
The control group receives the console output and the detailed log files produced by the respective agents.
Those participants can use a file viewer of their choice to analyze the logs, for which all participants chose the VSCode IDE.

\begin{table}[t]
\centering
\caption{Participants of the user study and assignment to groups.}
\label{tab:participants}
\small
\begin{tabular}{@{}llrr@{}}
\toprule
\multicolumn{2}{@{}l@{}}{Participant background} & Control group & Debugger group \\
\midrule
Occupation
& Bachelor's Computer Science Student & 3 & 2 \\
& Master's Computer Science Student & 0 & 1 \\
& PhD Candidate in Computer Science & 3 & 3 \\
\midrule
Programming experience
& 1-3 years & 1 & 2 \\
& 3-5 years & 3 & 2 \\
& 5-10 years & 2 & 1 \\
& 10+ years & 0 & 1 \\
\midrule
LLM familiarity
& None & 2 & 2 \\
& Basic (heard of them) & 2 & 1 \\
& Moderate (configured and used) & 1 & 1 \\
& Advanced (developed or analyzed) & 1 & 2 \\
\bottomrule
\end{tabular}
\end{table}

Table~\ref{tab:participants} shows the assignment of participants to the two groups.
Both groups contain three computer science students and three PhD candidates.
Each group includes two participants with no prior knowledge of LLM agents and two to three participants with moderate or advanced knowledge.
Both groups also have similar programming experience.

Once assigned to their respective groups, each participant completes all three tasks.
That is, the total number of task instances completed in the study is 36 (12 participants $\times$ 3 tasks).

\paragraph{Procedure}

Each participant books a one-hour session in advance.
To ensure consistency, all sessions follow the same structure and are conducted by the same experimenter.
The experimenter follows a standardized study script and reads it verbatim, providing identical information to all participants.
Participants receive a three-minute introduction to LLM agents, covering high-level background information and the typical structure and workflow of agents.
After this introduction, participants are informed of their group assignment and receive a three-minute explanation of the tools available for completing the tasks. Additionally, participants have three minutes to familiarize themselves with the tools.
Following this orientation, participants complete the tasks. Each task is timed.

\subsubsection{Results}

\begin{figure}
  \includegraphics[width=\linewidth]{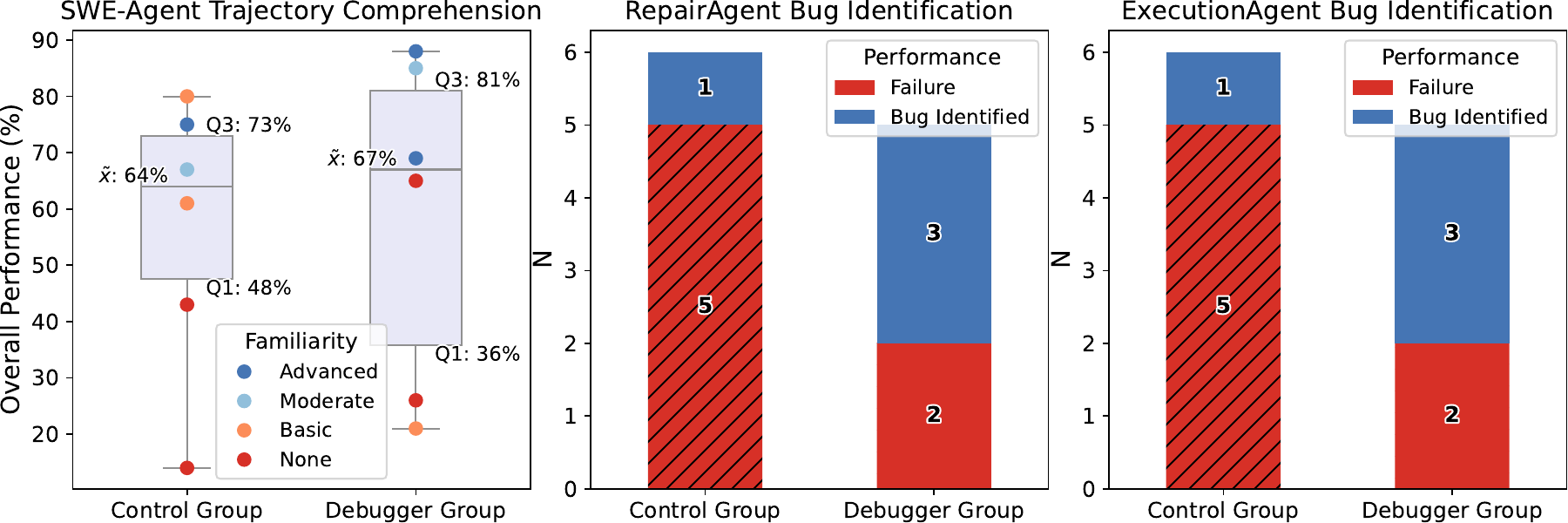}
  \caption{RQ2 results for trajectory comprehension task (left) and the two bug localization tasks (middle and right).}
  \label{fig:task_performance}
\end{figure}

\paragraph{Trajectory comprehension task}

Figure~\ref{fig:task_performance} (left) presents the results of the trajectory comprehension task.
The plot shows the scores achieved by the participants in both groups, where 100\% corresponds to a perfect score.
The median performance achieved by the control group and debugger group is 64\% and 67\%, respectively.
To illustrate the performance of individual participants, the dots in the plot represent the scores of each participant, which we color depending on the familiarity of the participant with LLM agents.
For both groups, we observe a high variance, with participants who are more familiar with LLM agents generally achieving higher scores, whereas participants with little or no familiarity tend to achieve lower scores.
Overall, the participants in the debugger group slightly outperform those in the control group, with a difference in median performance of 3 percentage points.

\paragraph{Bug identification tasks}

The middle and right-hand plots in Figure~\ref{fig:task_performance} present the results of the two bug identification tasks.
For each task and group, the plot shows how many of the participants correctly identified the bug in the agent program (blue) or failed to do so (red, shaded).
The debugger group has only five participants for the bug identification tasks, as we remove one participant who has prior experience with the two agents used in these tasks to avoid bias.
As shown in the middle plot, only one of the control group participants correctly identifies the bug in RepairAgent, whereas three out of five participants in the debugger group do so.
Similarly, in the right-hand plot, only one participant in the control group correctly identifies the bug in ExecutionAgent, whereas three participants in the debugger group do so.
While the total numbers are the same for the two tasks, the individual participants who correctly identify the bugs differ between the tasks.
Overall, the debugger group clearly outperforms the control group, with a total of $2/12=17\%$ of participants in the control group correctly identifying the bugs, compared to $6/10=60\%$ in the debugger group.

\paragraph{Discussion}

To statistically analyze the results, we use the Mann-Whitney U~test~\cite{MannWhitneyUTest} with a significance level of $p < 0.05$ to compare the two groups.
Due to the small sample size and diverse participant backgrounds, detecting statistically significant differences is challenging.
Nevertheless, we observe a statistically significant difference for two out of the three studied tasks: For the two bug localization tasks, the debugger group outperforms the control group in a statistically significant manner.

Taking a more qualitative perspective, we observe two main reasons why \name{} supports users in understanding and debugging software development agents.
One reason is that the approach enables users to see information that otherwise is very hard to obtain from the logs alone.
For example, the bug in RepairAgent causes a code change to be not applied to the code base, despite the corresponding tool call returning a success message.
The fact that \name{} captures intermediate code changes made by the agent and presents them to the user as diffs allows users to quickly identify this discrepancy.
The other reason is that \name{} allows users to retrieve information about an agent's behavior more easily than by inspecting log files.
In particular, the single-sentence summary of each event helps users to quickly grasp the essence of an event without having to read through lengthy log messages.

\subsection{RQ3: Workload Perceived by Agent Developers}

To answer the question of how using \name{} affects the perceived workload of developers, we ask participants to assess the subjective difficulty of performing the tasks.
After each task, participants complete the NASA Task Load Index (TLX)~\cite{NASATLX} form, a standardized questionnaire that measures perceived workload across multiple dimensions: mental demand, temporal demand, effort, performance, and frustration.\footnote{We omit the ``physical demand'' dimension, as it does not apply to computer-based tasks.}

\begin{figure}
  \includegraphics[width=.9\linewidth]{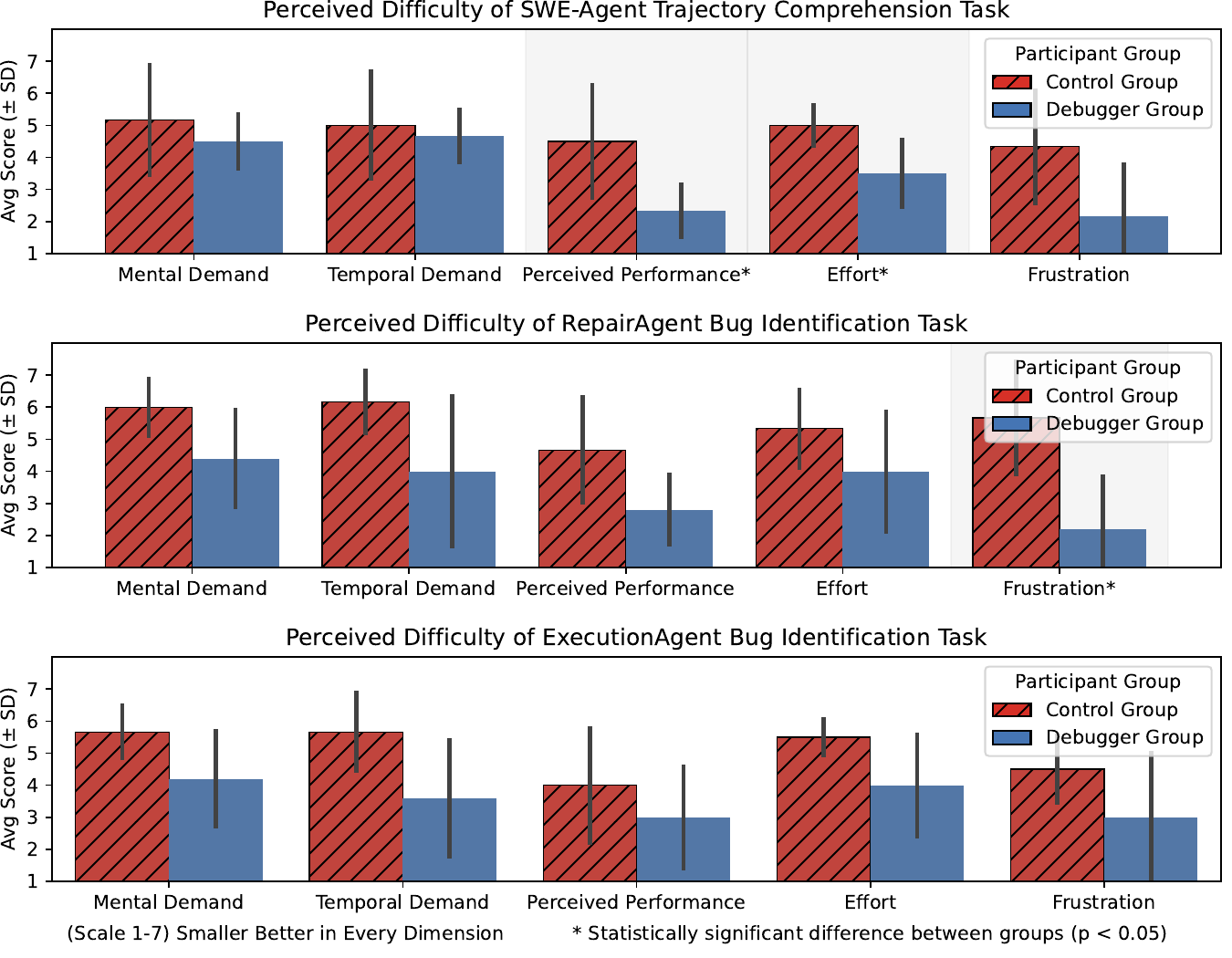}
  \caption{RQ3 results showing NASA TLX scores for the three tasks (lower = better).}
  \label{fig:task_tlx}
\end{figure}


Figure~\ref{fig:task_tlx} shows the NASA TLX scores for the three tasks, where a lower score indicates a lower perceived workload.
We find that participants in the debugger group consistently report lower workload across all three tasks compared to the control group.
The differences are most pronounced for the perceived performance, i.e., how successful participants feel in accomplishing the tasks, and for frustration, i.e., how insecure, discouraged, irritated, and annoyed participants feel during the tasks.
For example, the average frustration score across all three tasks is 5.4 for the control group, whereas it is only 2.4 for the debugger group.
We apply the Mann-Whitney U~test~\cite{MannWhitneyUTest} with a significance level of $p < 0.05$ and mark statistically significant differences in the figure with an asterisk.
Similar to RQ2, the small sample size yields limited statistical power, but we observe statistically significant differences for three of the individual task-dimension combinations: perceived performance and effort on the trajectory comprehension task, and frustration on the RepairAgent bug identification task.
Beyond statistical significance, the consistent trend of lower workload scores in the debugger group across all tasks and dimensions suggests that \name{} effectively supports users in understanding and debugging software development agents.

\subsection{Threats to Validity}
Our evaluation has certain limitations that may affect its validity and generalizability.
First, the code changes and API calls metrics for RQ1 may not fully capture all integration effort, such as understanding agent architecture.
We select these metrics because they provide objective, quantifiable measures enabling consistent comparison.
Second, the paper's main author was not involved in the original agents' development; the original developers users are likely to integrate with comparable or less effort given their familiarity.
Third, the user study involves a relatively small participant group, limiting statistical power.
We mitigate this by carefully balancing group assignments based on technical background.
Fourth, study tasks may not fully represent real-world debugging scenarios.
We address this by using multiple activity types and three agents.
Fifth, participants were not original developers, potentially affecting their performance.
Results suggest participants with more LLM agent familiarity benefit from \name{} at least as much as those with less familiarity.
Finally, focusing on three specific agents limits generalizability.
We mitigate this by selecting agents differing in tasks, architectures, and toolsets.

\section{Related Work}

\paragraph{Software engineering agents}

Recent years have seen a surge of interest in software engineering agents that leverage LLMs to assist developers in various tasks.
Program repair and issue solving are the most prominent tasks, with notable examples including
RepairAgent~\cite{icse2025-RepairAgent},
AutoCodeRover~\cite{Zhang2024a},
OpenHands~\cite{Wang2024a},
SWE-Agent~\cite{Yang2024a},
Magis~\cite{Tao2024},
AgentCoder~\cite{Huang2024},
MarsCode Agent~\cite{Liu2024a},
FixAgent~\cite{Lee2024}, and
Trae Agent~\cite{gao2025trae}.
Other tasks addressed by software engineering agents include
environment setup, e.g., by ExecutionAgent~\cite{issta2025_ExecutionAgent} and others~\cite{Milliken2025,Eliseeva2025,Yang2025,hu2025llm},
root cause analysis~\cite{Roy2024},
generating issue-reproducing tests~\cite{Muendler2024,Ahmed2024,icse2026_Issue2Test,Cheng2025}, and 
debugging computational notebooks~\cite{Grotov2024}.
Rather than proposing a new agent, our work provides a technique to improve the development of existing and future software engineering agents by enabling interactive debugging.
Trust has been identified as a key concern in software engineering agents~\cite{cacm2025_agenticSE}.
We see interactive debugging as one way to increase trust in an agent's behavior.

\paragraph{Studies of software engineering agents}

To better understand the behavior of software engineering agents, several recent works have conducted empirical studies~\cite{ase2025_agent_study,Ceka2025}.
These studies yield insights into agent performance, common failure modes, and typical interaction patterns.
Our work could support such studies by providing a tool for interactively exploring and analyzing agent trajectories.
Others aim to automatically identify error types in agent trajectories~\cite{deshpande2025trailtracereasoningagentic}, which is orthogonal to our goal of enabling human developers to debug agents.
Finally, some studies are primarily about the end-to-end effectiveness of software engineering agents~\cite{Rondon2025}, which differs from our goal of debugging the internal behavior of an agent.

\paragraph{Debugging LLM agents}

Most closely related to our work on tools and techniques for debugging LLM agents.
Several of them focus on multi-agent systems, enabling inspection of the messages exchanged between multiple agents~\cite{Epperson2025}, their social interaction patterns~\cite{lu2024agentlens}, or trying to attribute failures to specific agents via spectrum-based fault localization~\cite{ge2025introducingfailureautomaticallyattributing}.
%
%
%
%
Raggy~\cite{lauro2025raglaginteractivedebugging} offers interactive debugging tool for retrieval-augmented generation (RAG) systems, helping to understand the impact of hyperparameters, such as the number of retrieved documents.
Unlike \name{}, none of the above techniques is designed for LLM agents that interact with external tools, and none offer specific support for software development agents, such as displaying code changes made by an agent.
Watson~\cite{rombaut2025watsoncognitiveobservabilityframework} proposes a surrogate agent that operates in parallel with the main agent, trying to reach the same result while also providing explanations for its actions.
Finally, some commercial providers of LLMs offer tools to inspect trajectories:
e.g., OpenAI's Platform Tools offers a tracing dashboard that displays LLM calls and responses made by an agent in real time.
However, it is not interactive, cannot pause or modify agent execution, and also offers no specific support for agents that interact with code repositories.

\paragraph{Debugging}

Motivated by the never-ending imperfection of software, debugging traditional software has a long history.
Popular techniques include text-based, interactive debuggers, such as the GNU debugger (gdb)~\cite{stallman1988debugging},
back-in-time debugging~\cite{Lienhard2008},
question-based debugging~\cite{Ko2008},
statistical debugging~\cite{Liblit2005,Chilimbi2009},
performance debugging~\cite{Han2012,Song2014},
and delta debugging, which identifies those parts of a program input responsible for a failure~\cite{dd,hdd,ase2017-GTR}.
Our work builds on this rich history by adapting the concept of interactive debugging to the context of software development agents, which present unique challenges due to their reliance on LLMs and interactions with external tools.
Other work explores how to enable LLMs to interact with debuggers, either to partially automate the debugging process~\cite{Levin2024,Bajpai2024} or to improve LLM-based code generation~\cite{Zhong2024,Yuan2025}.
While that line of work lets LLMs use debuggers, our work focuses on enabling human developers to debug LLM-based software development agents.

\section{Conclusion}

This paper presents \name{}, the first interactive debugger for software development agents that enables developers to step through an agent's execution, inspect its internal state, and modify its behavior at runtime.
Our evaluation demonstrates that integrating \name{} into existing agents requires only modest effort, and a user study shows that \name{} helps developers better understand and debug software development agents while reducing their perceived workload.
We envision our approach to contribute to the development of more reliable and trustworthy software development agents, which is an essential step toward their broader adoption in practice. 

\section{Data Availability}

Our implementation and all relevant data are available at:\\
\url{https://github.com/sola-st/AgentStepper} 

\bibliographystyle{ACM-Reference-Format}
\bibliography{referencesMichael,referencesMore}

\end{document}